\documentclass[journal]{IEEEtran}

\usepackage{cite}
\usepackage[pdftex]{graphicx}
\usepackage{amsmath}
\usepackage{color}
\newcommand{\red}[1]{{\color{black}#1}}
\usepackage{amsmath,amssymb}
\usepackage{microtype}
\usepackage{epstopdf}

\begin{document}

\title{On Antenna Array Out-of-Band Emissions}

\author{Lauri~Anttila,~\IEEEmembership{Member,~IEEE,}
        Alberto~Brihuega,~\IEEEmembership{Student Member,~IEEE,}
        Mikko~Valkama,~\IEEEmembership{Senior~Member,~IEEE}
\thanks{The authors are with the Department of Electrical Engineering, Tampere University, Tampere,
Finland; e-mail: lauri.anttila@tuni.fi.}}
\vspace{-15mm}

\maketitle

\begin{abstract}
We substantiate and extend recent research on the behavior of the out-of-band emissions in antenna array transmitters.  
Specifically, with multi-user precoding, we show that the emissions are always beamformed in the directions of the intended receivers, contrary to some claims in the recent literature. Moreover, while also other spurious directions exist, we show that the emissions are always strongest in the directions of the intended receivers. We also show that power amplifiers with mutually different nonlinear phase characteristics will reduce the beamforming gain of the unwanted emissions, with the gain approaching the noncoherent combining limit as the phase deviations are increased.
\end{abstract}


\section{Introduction}
\IEEEPARstart{P}{ower} amplifiers (PA) in radio transmitters are always nonlinear. Such nonlinearities induce both in-band and out-of-band (OOB) emissions, which may limit the capacity and quality of service of the intended and neighboring channel users. The transmit signal quality and unwanted emissions are governed by radio standardization bodies, typically through error vector magnitude (EVM), adjacent channel leakage ratio (ACLR), and spurious emission limits. The problem is well known in traditional single and few-antenna systems, while the behavior of the unwanted emissions in large antenna arrays, especially with multiple users, has been studied less thoroughly, though research on this topic is mounting \cite{Larsson_OOB_2018,Mollen_OOB_2018,Zou_2015,S_Blandino}.

In \cite{Mollen_OOB_2018}, a rigorous analysis based on Hermite-orthogonal polynomials for representing the amplified signal was conducted. By leveraging the spatial cross-correlation matrix of the nonlinear distortion, the behaviour of the OOB emissions was exposed under different propagation scenarios in single-user and multi-user transmission. However, the analysis was conducted considering an identical polynomial model across the array. In this letter, we consider a more general case with mutually different polynomial models for the parallel PAs and expound on the potential implications of such differences. Additionally, in \cite{S_Blandino}, analysis similar to \cite{Mollen_OOB_2018} was conducted based on the spatial cross-correlation matrix of the distortion. Different from \cite{Mollen_OOB_2018}, it was claimed  that the array gain is only experienced within the desired band, implying that the nonlinear distortion would not combine coherently. However, this claim has been refuted in many other works \cite{Larsson_OOB_2018,Zou_2015,Mollen_OOB_2018,Digital_MIMO_DPD,Mahmoud_Subarray_DPD}, where the nonlinear distortion has been shown to follow essentially the same spatial response as the inband signal.

Recently, in \cite{Larsson_OOB_2018}, the authors used a simple two-tone signal model to argue that the PA-induced OOB emissions of antenna arrays, in a multi-user scenario, are "beamformed into distinct spatial directions, which are different from those of the desired signals." In this letter, we adopt a more general I/Q modulated signal model and show that this conclusion is largely incomplete. Particularly, we show that some of the \red{inband and} OOB emissions are \textit{always} beamformed in the same directions as the desired signals. Moreover, we also show that the emissions are always strongest in these directions. Finally, we quantitatively address the impact of the phase deviations in the nonlinear behavior of the involved PA units.

\section{Signal Models}\label{Modelling}

\subsection{Basic Two-tone Model}

\begin{figure}[!t]
  \begin{center}
    \includegraphics[width=0.9\columnwidth]{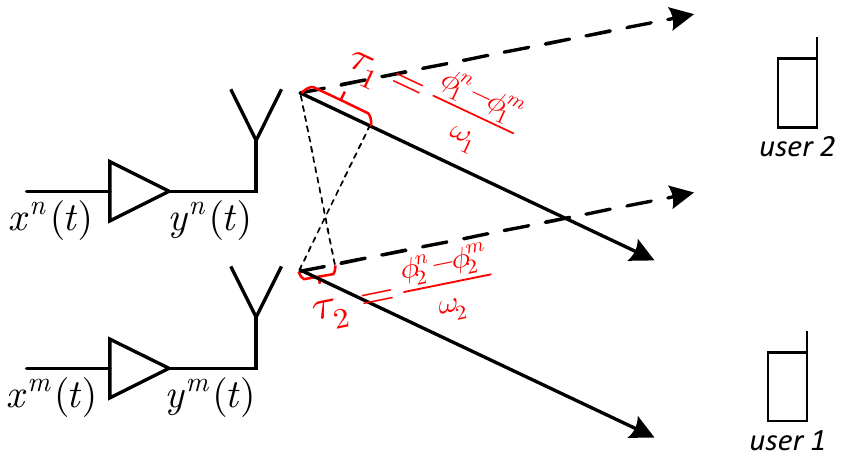}
  \end{center}
  \caption{Relation between the beamforming phases and the propagation delays between two antennas in a two-user system.}
  \label{fig:tau}
\end{figure}

To show the limitations of the results in \cite{Larsson_OOB_2018}, we start with a basic two-tone test signal written for the $m$th PA input 
\begin{align}
    x^m(t) = \cos(\omega_1 t+\phi_1^m) + \cos(\omega_2 t+\phi_2^m),
\end{align}
where $\omega_{1}$ and $\omega_{2}$ are the tone (angular) frequencies, and $\phi_{1}^{m}$ and $\phi_{2}^{m}$ are the phases that control the beamforming direction. Fig. \ref{fig:tau} illustrates how the beamforming phases are related to the propagation delays $\tau_1, \tau_2$ and the carrier frequencies $\omega_1, \omega_2$ between two antennas in a two-user system. As in \cite{Larsson_OOB_2018}, the PAs are modeled with a strictly memoryless 3rd-order polynomial
\begin{align}
    f(x)=x+\alpha x^3,
\end{align}
which yields the PA output signal of the form \cite{Larsson_OOB_2018} 
\begin{align}
    y^m(t)&= \Big(1+\frac{9 \alpha}{4} \Big) \cos(\omega_1 t+\phi_1^m) \label{eq:d1} \\
    &+ \Big(1+\frac{9 \alpha}{4} \Big) \cos(\omega_2 t+\phi_2^m) \label{eq:d2} \\
    &+ \frac{3 \alpha}{4} \cos(2\omega_2 t - \omega_1 t + 2\phi_2^m - \phi_1^m) \label{eq:im1} \\
    &+ \frac{3 \alpha}{4} \cos(\omega_2 t - 2\omega_1 t + \phi_2^m - 2\phi_1^m), \label{eq:im2}
\end{align}
where the intermodulations that are far away from the allocated band have been omitted for brevity.
Referring to the definitions in Fig. \ref{fig:tau}, it is easy to see that (\ref{eq:im1}) and (\ref{eq:im2}) add up coherently in the respective directions defined by 
\begin{align}
  \tilde{\tau}_1 &= \frac{(2\phi_2^n - \phi_1^n) - (2\phi_2^m - \phi_1^m)}{2\omega_2-\omega_1} \label{eq:tau1} \\
  \tilde{\tau}_2 &= \frac{(\phi_2^n - 2\phi_1^n) - (\phi_2^m - 2\phi_1^m)}{\omega_2-2\omega_1}. \label{eq:tau2}
\end{align}

It was stated in \cite{Larsson_OOB_2018}, while referring to (\ref{eq:d1}) and (\ref{eq:d2}), that "the first two terms represent the desired signal, scaled by a constant," and were omitted in the subsequent distortion analysis. While this is a correct interpretation in the most simple single-user/two-tone signal case, we show in the following  
that it is incorrect in the general multi-user/modulated signal case.
\subsection{Narrowband I/Q Modulated Signal Model}
Consider now a general two-carrier I/Q modulated signal as the PA input, defined as
\begin{align}
    x^m(t) &= A_1(t)\cos(\omega_1 t + \theta_1(t) + \phi_1^m) \nonumber \\
    &+ A_2(t)\cos(\omega_2 t + \theta_2(t) + \phi_2^m).
\end{align}
The corresponding complex baseband signals are denoted as $s_1(t)=A_1(t) \text{exp}(j \theta_1(t))$ and $s_2(t)=A_2(t) \text{exp}(j \theta_2(t))$, and we assume that the typical narrowband assumption applies to them, i.e., phase steering works over the full bandwidth of the signals \cite{Mailloux_1982}.

Now, repeating the derivation for the PA output signal with the I/Q modulated input signal, we obtain

\begin{align}
    y^m (t)
    &= \Big( A_1 + \frac{3 \alpha}{2} A_1 A_2^2 + \frac{3 \alpha}{4} A_1^3 \Big) \cos(\omega_1 (t)+\phi_1^m) \\
    &+ \Big( A_2 + \frac{3 \alpha}{2} A_1^2 A_2 + \frac{3 \alpha}{4} A_2^3 \Big)  \cos(\omega_2 (t)+\phi_2^m) \\
    &+ \frac{3 \alpha}{4} A_1 A_2^2 \cos(2\omega_2 (t) - \omega_1 (t) + 2\phi_2^m - \phi_1^m) \\
    &+ \frac{3 \alpha}{4} A_1^2 A_2 \cos(\omega_2 (t) - 2\omega_1 (t) + \phi_2^m - 2\phi_1^m),
\end{align}
where again the intermodulations that are far away from the allocated band have been omitted. Here, in order to have a compact expression, we have used the shorthand notation $\omega_l (t) = \omega_l t+\theta_l(t)$ and dropped the time argument from the amplitude signals.

The above expression now reveals a clearly more complex structure for the first two terms, compared to (3) and (4). They do not represent only the desired signal, but also contain intermodulation and crossmodulation distortions from the two amplitude signals. Thus, the complete list of relevant distortion terms near the frequencies $\omega_1$ and $\omega_2$ is given by

\begin{align}
    &\Big( \frac{3 \alpha}{2} A_1 A_2^2 + \frac{3 \alpha}{4} A_1^3 \Big) \cos(\omega_1 (t)+\phi_1^m) \label{eq:dist1} \\
    &\Big( \frac{3 \alpha}{2} A_1^2 A_2 + \frac{3 \alpha}{4} A_2^3 \Big)  \cos(\omega_2 (t)+\phi_2^m) \label{eq:dist2} \\
    &\frac{3 \alpha}{4} A_1 A_2^2 \cos(2\omega_2 (t) - \omega_1 (t) + 2\phi_2^m - \phi_1^m) \label{eq:dist3} \\
    &\frac{3 \alpha}{4} A_1^2 A_2 \cos(\omega_2 (t) - 2\omega_1 (t) + \phi_2^m - 2\phi_1^m). \label{eq:dist4}
\end{align}

 Considering then true spatial multiplexing, i.e., $\omega_1 = \omega_2 = \omega$, and substituting back $\omega_l (t) \leftarrow \omega t+\theta_l(t)$, these reduce to
\begin{align}
    & \Big( \frac{3 \alpha}{2} A_1 A_2^2 + \frac{3 \alpha}{4} A_1^3 \Big) \cos(\omega t+\theta_1(t)+\phi_1^m) \label{eq:z1} \\
    & \Big( \frac{3 \alpha}{2} A_1^2 A_2 + \frac{3 \alpha}{4} A_2^3 \Big)  \cos(\omega t+\theta_2(t)+\phi_2^m) \label{eq:z2} \\
    & \frac{3 \alpha}{4} A_1 A_2^2 \cos(\omega t + 2\theta_2(t) - \theta_1(t) + 2\phi_2^m - \phi_1^m) \label{eq:u1} \\
    & \frac{3 \alpha}{4} A_1^2 A_2 \cos(\omega t - \theta_2(t) + 2\theta_1(t) - \phi_2^m + 2\phi_1^m). \label{eq:u2}
\end{align}
The first two distortion terms add up coherently in the directions of the corresponding intended receivers, while the two latter terms are beamformed into directions defined by (\ref{eq:tau1}) and (\ref{eq:tau2}). \red{We note that in general, the spectral contents of the above distortion terms are covering both the inband and OOB frequencies.}

\section{Beamformed Distortion Power Analysis}

We next analyze the powers of the distortion terms (\ref{eq:z1})-(\ref{eq:u2}) from the beamformed channel perspective, utilizing their baseband equivalent forms. For generality, we also consider and allow for mutually different power amplifiers, with complex-valued coefficients of the form $\alpha_m = \bar{\alpha} \cdot \text{exp}(j\psi_m)$. 
Here, $\bar{\alpha}$ denotes the baseband equivalent PA coefficient corresponding to the real-valued PA coefficient $\alpha$, and $\psi_m$ are random, zero-mean, phase deviations. While the distortion coefficients of true PA's may also exhibit amplitude differences between the units, we focus specifically on the phase variations, because they are more critical from the beamforming perspective.

\subsection{Observed Baseband Equivalent Distortion Terms}
Now, since the beamforming coefficients are chosen such that coherent combining takes place in the directions defined by $\tau_1$ and $\tau_2$, the equivalent beamformed distortions observed by the intended users are given by 

\begin{equation}
\begin{split}
    r_1(t) &= \sum_{\substack{m=1}}^{M}\Big( \frac{3\alpha_m}{2} A_1(t) A_2^2(t) + \frac{3\alpha_m}{4} A_1^3(t) \Big)e^{j\theta_1(t)}\\
           &= z_{1}(t) \sum_{\substack{m=1}}^{M}e^{j\psi_m}\label{eq:r1}
\end{split}
\end{equation}
\begin{equation}
\begin{split}
    r_2(t) &= \sum_{\substack{m=1}}^{M}\Big( \frac{3\alpha_m}{2} A_1^2(t) A_2(t) + \frac{3\alpha_m}{4} A_2^3(t) \Big)e^{j\theta_2(t)}\\
           &= z_{2}(t)\sum_{\substack{m=1}}^{M}e^{j\psi_m}\label{eq:r2},
\end{split}
\end{equation}
where $z_{1}(t) = (\frac{3}{2}\bar{\alpha} A_1(t) A_2^2(t) + \frac{3}{4}\bar{\alpha} A_1^3(t))e^{j\theta_1(t)}$ and $z_{2}(t) = (\frac{3}{2}\bar{\alpha} A_1(t)^2 A_2(t) + \frac{3}{4}\bar{\alpha} A_2^3(t))e^{j\theta_2(t)}$. 

On the other hand, potential victim receivers located in the directions defined by the delays $\tilde{\tau}_1$ and $\tilde{\tau}_2$ would observe effective beamformed distortions of the form

\begin{equation}
\begin{split}
    v_1(t) &= \sum_{\substack{m=1}}^{M}\Big( \frac{3\alpha_m}{4} A_1(t) A_2^2(t) \Big)e^{j(2\theta_2(t) - \theta_1(t))}\\
           &= u_{1}(t) \sum_{\substack{m=1}}^{M}e^{j\psi_m} \label{eq:v1}
\end{split}
\end{equation}
\begin{equation}
\begin{split}
    v_2(t) &= \sum_{\substack{m=1}}^{M}\Big( \frac{3\alpha_m}{4} A_1^2(t) A_2(t) \Big)e^{j( - \theta_2(t) + 2\theta_1(t))}\\
           &= u_{2}(t) \sum_{\substack{m=1}}^{M}e^{j\psi_m}\label{eq:v2},
\end{split}
\end{equation}
where $u_{1}(t) = \frac{3}{4}\alpha A_1(t) A_2^2(t) e^{j(2\theta_2(t) - \theta_1(t))}$ and $u_{2}(t) = \frac{3}{4}\alpha A_1^2(t) A_2(t) e^{j(2\theta_1(t) - \theta_2(t))}$. The term $\sum_{\substack{m=1}}^{M}e^{j\psi_m}$ in (\ref{eq:r1})-(\ref{eq:v2})  
can be interpreted as the \emph{effective beamforming gain}. 

\subsection{Power Analysis}
To evaluate the powers of the distortion terms in (\ref{eq:r1})-(\ref{eq:v2}), we assume that the baseband waveforms $s_1(t)$ and $s_2(t)$ are mutually independent complex-circular Gaussian signals with zero mean and variances $P_{s,l}=\text{E}[\left| s_l(t) \right|^2]$, while $\psi_m$  is assumed to be either Gaussian distributed $\psi_m \sim  \mathcal{N}(0,\,\sigma^{2})$ or uniformly distributed  $\psi_m\sim \mathcal{U}(-\sigma,\sigma)$. \red{We note that in this analysis, we address the total powers of the different terms while pursue the differentiation between the inband and OOB distortion powers along the numerical results.}

The powers of the distortion terms  (\ref{eq:r1})-(\ref{eq:r2}), which are received by the intended receivers, are given by
\begin{equation}
\begin{split}
    P_{r,l} &= \mathbb{E} \left\{\Big| z_{l}(t)\sum_{\substack{m=1}}^{M}e^{j\psi_m}\Big| ^2\right\} \\
          &=  \mathbb{E} \left\{\Big|z_{l}(t)\Big| ^2\right\}\mathbb{E} \left\{\Big|\sum_{\substack{m=1}}^{M}e^{j\psi_m}\Big| ^2\right\}\label{eq:beamformed_power}.
    \end{split}
\end{equation}
First, evaluating the left-hand side expectation operation, which represents the power of the baseband equivalents of (\ref{eq:z1})-(\ref{eq:z2}), we obtain
\begin{align}
    P_{z,l} = \frac{27 \alpha_m^2}{8} P_{s,l}^3 + \frac{9 \alpha_m^2}{2} P_{s,1}P_{s,2}^2 + \frac{9 \alpha_m^2}{2} P_{s,1}^2P_{s,2}, 
\end{align}
where the identities $\text{E}[\left| s_l(t) \right|^4]=2P_{s,l}^2$ and $\text{E}[\left| s_l(t) \right|^6]=6P_{s,l}^3$ \cite{Picinbono_1993} have been utilized. With equal-power user signals, $P_{s,1}=P_{s,2}=P_s$, these reduce to 
\begin{align}
    P_{z,1}=P_{z,2}=\frac{99 \alpha_m^2}{8} P_s^3.
\end{align}

The right-hand side expectation in (\ref{eq:beamformed_power}) represents the effective beamforming power gain. Considering first that $\psi_m \sim  \mathcal{N}(0,\,\sigma^{2})$, this yields
\begin{equation}
    G_{b}^{\mathcal{N}} =\mathbb{E} \left\{\Big| \sum_{\substack{m=1}}^{M}e^{j\psi_m}\Big| ^2\right\} =M +(M^2-M)e^{-\sigma^2},\label{normal_bf_gain}
\end{equation}
which follows from applying ordinary trigonometric identities and the following integral identity \cite{Integral}
\begin{equation}
    \frac{1}{\sqrt{2\pi\sigma^2}}\int_{-\infty}^{+\infty} \mathrm{cos}\psi_me^{-\frac{-\psi^2_m}{2\sigma^2}} d\psi_m =  e^{-\sigma^2/2}.
\end{equation}

On the other hand, if $\psi_m$ is uniformly distributed in the interval $\psi_m\sim \mathcal{U}(-\sigma,\sigma)$, the effective beamforming power gain can be shown to read
\begin{equation}
    G_{b}^{\mathcal{U}} =\mathbb{E} \left\{\Big| \sum_{\substack{m=1}}^{M}e^{j\psi_m}\Big| ^2\right\} =M +(M^2-M) \mathrm{sinc}^2\sigma, 
    \label{uniform_bf_gain}
\end{equation}
where $\mathrm{sinc}\sigma= \frac{\mathrm{sin}\sigma}{\sigma}$.
Consequently, the beamformed distortion towards the intended users, with equal-power signals, has a power of
\begin{align}
    P_r^{\mathcal{N}} &= \frac{99 \alpha_m^2}{8} P_s^3 (M+(M^2-M)e^{-\sigma^2}) \label{P_r_gaussian} \\
    P_r^{\mathcal{U}} &= \frac{99 \alpha_m^2}{8} P_s^3 (M+(M^2-M). \mathrm{sinc}^2\sigma) \label{P_r_uniform} 
\end{align}

Similarly, evaluating the powers of the spurious distortion terms in (\ref{eq:v1}) and (\ref{eq:v2}), we arrive at
\begin{align}
    P_{v,1} &= \frac{9 \alpha_m^2}{8} P_{s,1}P_{s,2}^2 \\
    P_{v,2} &= \frac{9 \alpha_m^2}{8} P_{s,1}^2P_{s,2}. \label{eq:Pu1}
\end{align}
With $P_{s,1}=P_{s,2}=P_s$, these reduce to 
\begin{align}
    P_{v,1}=P_{v,2}=\frac{9 \alpha_m^2}{8} P_s^3. \label{eq:Pu2}
\end{align}
Therefore, the power of the beamformed  distortions in the spurious directions is
\begin{align}
    P_v^{\mathcal{N}} &= \frac{9 \alpha_m^2}{8} P_s^3 (M+(M^2-M)e^{-\sigma^2}) \label{P_v_gaussian} \\ 
    P_v^{\mathcal{U}} &= \frac{9 \alpha_m^2}{8} P_s^3 (M+(M^2-M). \mathrm{sinc}^2\sigma) \label{P_v_uniform} 
\end{align}

Comparing (\ref{P_r_gaussian}) and (\ref{P_r_uniform}) with (\ref{P_v_gaussian}) and (\ref{P_v_uniform}), we see that with equal-power user signals, the distortion terms observed in the directions of the intended receivers are 11 times, or about $10.4$ dB's, stronger than in the spurious directions.

\section{Numerical Examples and Results} \label{sec:Simulations}

In this section, we assess and validate the above nonlinear distortion analysis with numerical simulations. 
First, we consider the simple $3$rd order polynomial model case  
where the adopted PA model is given as $f(x)=x-0.0368 x|x|^2$, which has been obtained by fitting a $3$rd order polynomial to a true measured PA response, while constraining the linear coefficient to one.  
Then, as a more practical case, we also consider $9$th order polynomial PA models with complex coefficients, obtained from RF measurements on a set of PAs. These models are of the form $f_m(x) =\sum_{\substack{p=1}}^{5} \alpha_{m,2p-1} x|x|^{2p-2}$, while are also clipped to reflect the true saturation levels of the measured PAs. Among the measured PAs, the maximum phase difference between the 3rd-order coefficients is 0.55~radians while the corresponding standard deviation is 0.1145~radians. The considered waveform is an LTE-like 20 MHz OFDM waveform sampled at 120 MHz, including windowing to improve the spectral containment. OFDM signals are known to have quasi-Gaussian distribution due to the central limit theorem \cite{Picinbono_1993}, and are thus comparable to the earlier analysis assumptions. 
We consider a MIMO transmitter with 60 antennas and PAs serving two users in the same frequency band under pure line-of-sight (LOS) propagation, and phase-only baseband precoder is utilized. 

\begin{figure}[!t]
  \centering
  \includegraphics[width=1\linewidth]{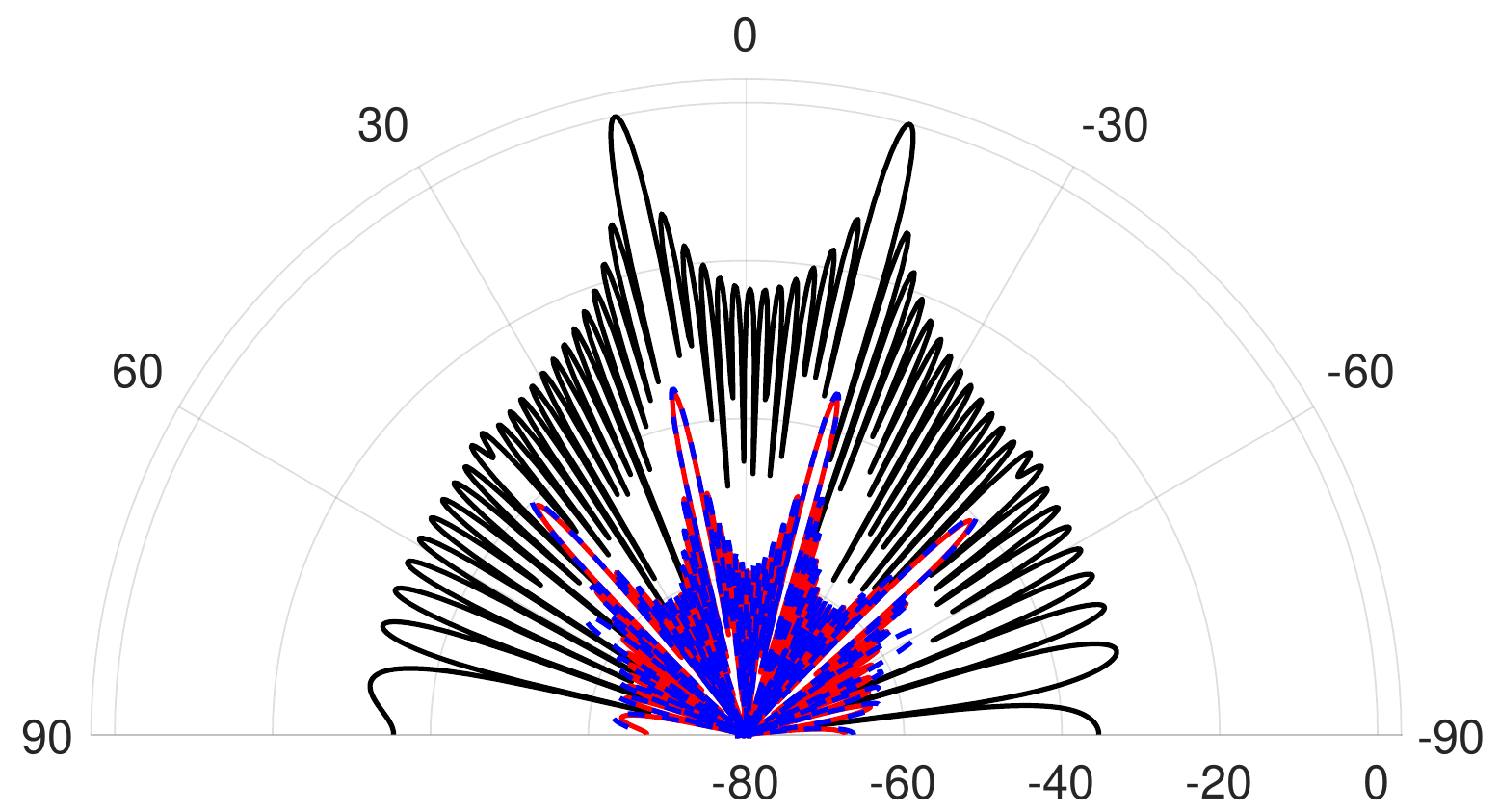}
  \caption[]{Full-bandwidth \red{total power based} beampatterns of the linear signal (dashed-line) and the four different distortion terms  
  stemming from the $3$rd order PA model, with 60 antenna units. The r-axis shows relative powers, with the passband power received by the intended receivers normalized to 0 dB.}
  \label{Fig.Polarplot1}
\end{figure}

Fig. \ref{Fig.Polarplot1} illustrates the corresponding beampatterns of the distortion terms in (\ref{eq:z1})-(\ref{eq:u2}) separately, along with the linear beampattern of the array (in dashed red line), with the main beams being steered towards $-15$ and $12$ degrees. The relative power of the distortion that gets beamformed towards the users' directions is $10.4$ dB stronger than that of the other spurious emissions (the distortions beamformed away from the main beam directions), which exactly matches the theory.

\begin{figure}[!t]
  \centering
  \includegraphics[width=1\linewidth]{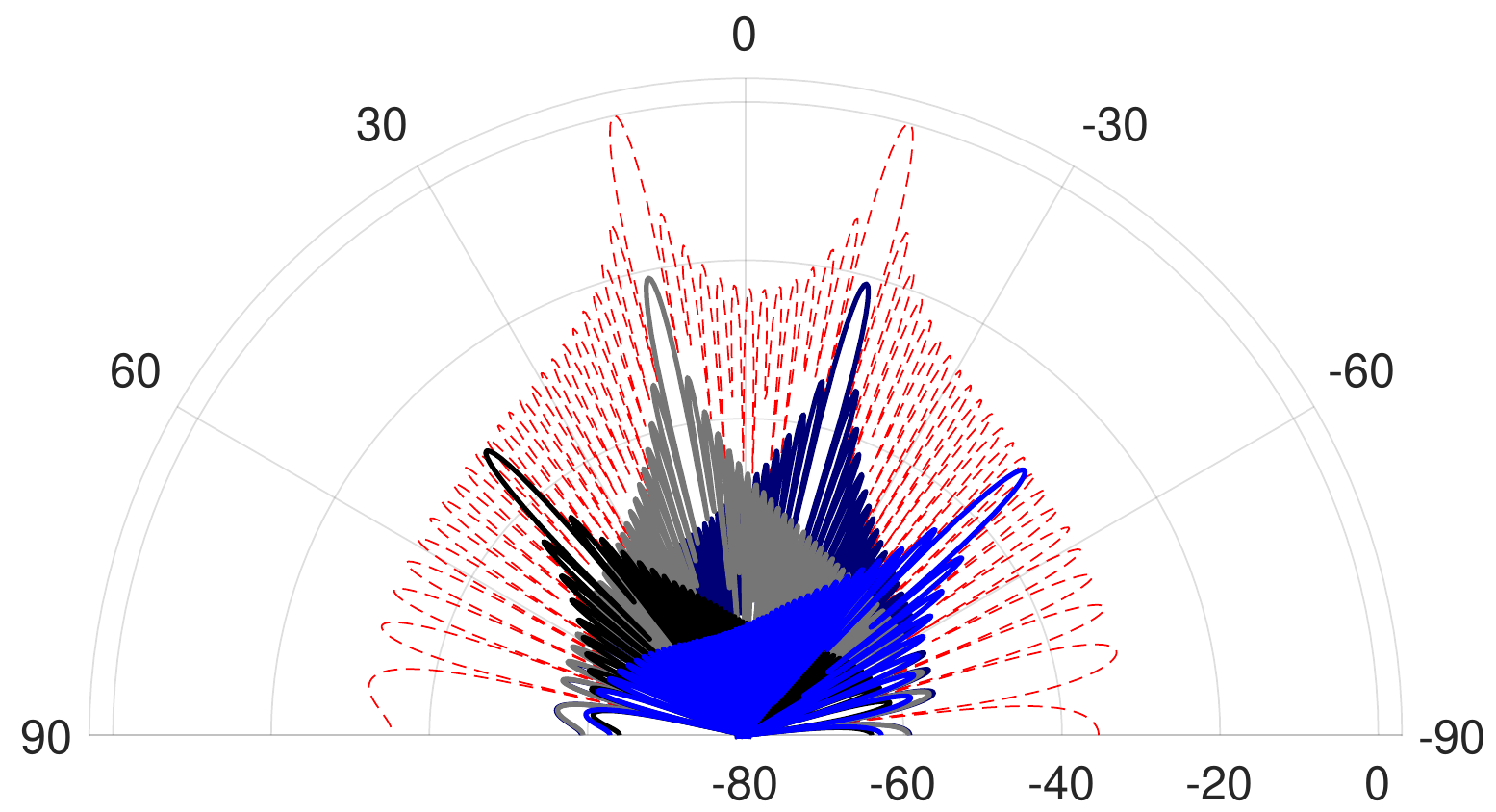}
  \caption[]{Inband power (black solid) and OOB emission power based beampatterns for the 3rd order (red solid) and the clipped 9th order (blue dashed) PA models with 60 antenna units. The passband power received by the intended receivers is normalized to 0 dB.}
  \label{Fig.Polarplot2}
\end{figure}

In Fig. \ref{Fig.Polarplot2}, we focus specifically on visualizing the spatial behaviour of the OOB distortion that is radiated from the transmitter, while also show the inband power based beampattern for reference. 
\red{The same time-domain signals are used, as in the previous example, while the inband and OOB distortion powers are extracted through FFT-based processing.}
As can be seen, the OOB distortion is beamformed partially towards the same directions as the passband and partially towards the directions defined by (\ref{eq:tau1}) and (\ref{eq:tau2}), exactly as predicted by (\ref{eq:dist1})-(\ref{eq:dist4}). 
This applies to both simple $3$rd order and more elaborate clipped $9$th order PA models, while in the $9$th order case there are also some other weaker spurious directions.  
These confirm the analytical results of Section II. 
In general, understanding the spatial behaviour of OOB distortions is key when developing efficient digital predistortion (DPD) solutions for multiple antenna transmitters \cite{Digital_MIMO_DPD,Mahmoud_Subarray_DPD}.

\begin{figure}[!t]
  \centering
  \includegraphics[width=.9\linewidth]{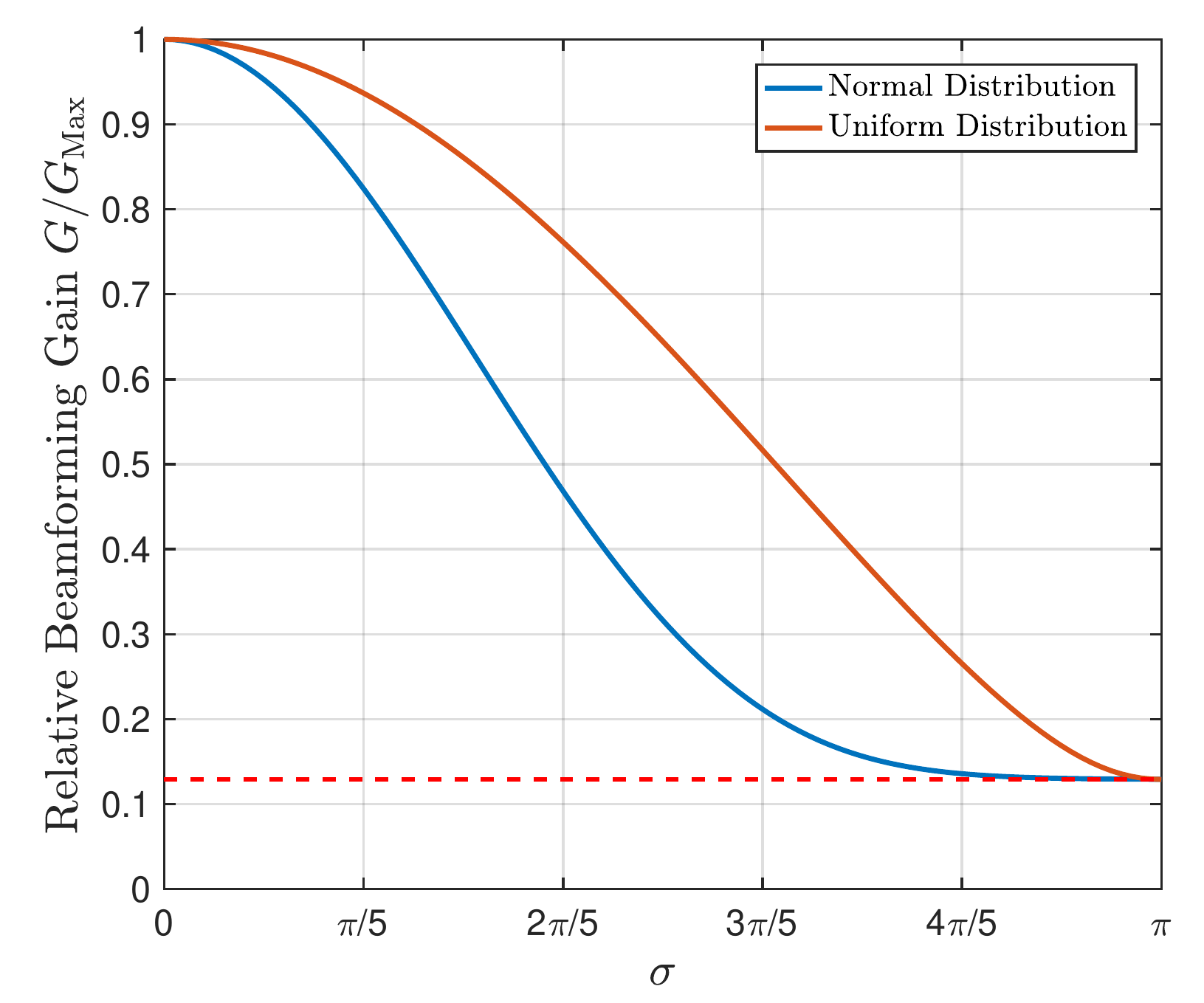}
  \caption[]{Relative beamforming gain as a function of $\sigma$ with $M=60$ antennas. The dashed line corresponds to the fully noncoherent combining gain of $\sqrt{M}$.}
  \label{Fig.Distributions}
\end{figure}

Lastly, Fig. \ref{Fig.Distributions} illustrates the relative beamforming gain of the distortion terms, given by the square root of (\ref{normal_bf_gain}) and (\ref{uniform_bf_gain}), and normalized by the maximum gain $M$. It can be observed, that the phase deviation of the 3rd order PA coefficients can have a significant impact on 
the amount of distortion that is observed by the different receivers. However, with small phase deviations, the impact is observed to be small.

\ifCLASSOPTIONcaptionsoff
  \newpage
\fi

\section{Conclusions }
In this letter, we analyzed the nonlinear distortions stemming from multi-user precoded large array transmitters with mutually different PAs. It was shown that the distortions are beamformed not only to spurious directions, but also to the main beam directions, contrary to some claims made in recent literature. Furthermore, the impact of differences in the PA phase characteristics on the beamformed distortion was analyzed, and they were shown to be potentially important in controlling the beamforming gain of the nonlinear distortions that are radiated from the transmitter. 
Proper understanding of the unwanted emissions in array transmitters is of fundamental importance for the development of array-based systems, and for developing efficient linearization solutions in them.

\bibliographystyle{IEEEbib}
\bibliography{ref}

\end{document}